# Sand Inclusion Composite Structures for Enhanced Ballistic Impact Resistance

Manas Thakur[*], Nakka Nishika[ŧ], Bommiditha Jyothsnavi[ŧ], Surkanti Sai Sahasra[ŧ], Srikant Sekhar Padhee[⫲]

[*]Dept. of Mechanical Engineering, MTech Student, Indian Institute of Technology Ropar, Rupnagar, India
[ŧ]Dept. of Mechanical Engineering, B.Tech. Student, Indian Institute of Technology Ropar, Rupnagar, India
[⫲]Dept. of Mechanical Engineering, Associate Professor, Indian Institute of Technology Ropar, Rupnagar, India

Email: *manasthakur159@gmail.com*
*nishika1028@gmail.com*
*jyothsnavi987@gmail.com*
*saisahasrasurkanti@gmail.com*
*sspadhee@iitrpr.ac.in*

## Abstract

With the ever-increasing threat of ballistic impact, it is essential to provide a solution that is not only effective but also economical. In recent years Polymer Matrix Sand Composites (PMSCs) have emerged as a potential cost-effective solution. In this ongoing effort, current investigation aims at offering more robust protection against varied ballistic impacts against potential ballistic threats. This research investigates the enhancement of ballistic impact resistance in polymer matrix composites through the inclusion of sand in a graded fashion. Variable material properties are obtained along the thickness by varying the sand inclusions size and weight fraction in the polymer matrix. The gradation offers a stepwise structure that has a prime base impact zone containing abrasive sand inclusions densely graded with a range of typical inclusion sizes which is brittle and hard enough to erode the projectile. Further gradation involves a less dense region providing tensile strength which is able to reflect the stress wave and reduce the impact energy with providing a backing to the frontal brittle region. The barely dense or neat matrix regions prevents backlash and offers cushioning effect. The addition of sand particles increases the surface area of the composite and improves adherence to the inclusions and matrix. This enhanced adhesion ensures efficient load transfer across the surface, which raises the composite's overall hardness. PMSCs were fabricated for observing the effects on two matrix compositions improvising manufacturing procedures and subjected to a battery of tests, including tensile testing, Izod impact testing, Shore-D hardness testing, etc. in order to evaluate their mechanical properties comprehensively. The experimental results reveal that varying the weight fraction of sand content markedly influences the mechanical properties up to a certain weight fraction for a typical size range of inclusions. Experimental and Simulation Studies were involved in order to extract the Mechanical Properties and to observe the material behavior against ballistic threats. The study reveals underscoring the potential of PMSCs as a cost-effective and environmentally sustainable Composites for advanced ballistic protection applications.

## 1. Introduction

Effective ballistic resistant structures often exhibit an amalgamation of several layers with either different material configurations or distinguished structural configurations which ultimately provides high hardness, toughness, brittle, lightweight, and high energy absorbent characteristics in the structures. Monolithic structures, which often consist of a single material like iron, steel, or aluminium alloys, are commonly used for ballistic-resistant structures. However, these structures tend to have high areal densities, resulting in excessive weight, making them inefficient in terms of weight management. Additionally, they usually exhibit low multi-hit capabilities, which reduces their effectiveness when subjected to repeated impacts. Moreover, the overall cost of such monolithic structures is often high, further limiting their suitability for ballistic impact applications [1-3]. Studies have been carried out in order to reduce the weight of the monolithic structure by incorporation of multi-layered structures such as sandwich structures comprising of regions with significant characteristics which combinedly work as the bullet pierces through the thickness of the structure. In order to restrict penetration, ballistic sandwich designs are usually constructed with a hard frontal layer, usually made of ceramics or Rolled Homogeneous Armor (RHA), to absorb initial impact energy. This causes bullet deformation and spreads contact stresses [4]. Energy-absorbent middle layers, such as composite metal foams (CMFs) or corrugated structures, are used to produce light weight structures and provide effective backing to frontal regions. Such arrangement helps mitigate catastrophic failure of the structure [5]. Further to distribute the impact forces and reduce the back face signature enabling impact damage mitigation, the last layer is usually made up of a thin ductile layer [6]. However, each material layer presents unique limitations. Brittle frontal layers, especially ceramics, are costly, heavy, and have limited multi-hit capacity, as they require substantial support to withstand fragmentation. Energy-absorbent cores, such as corrugated sheets and CMFs, face manufacturing challenges and limited multi-hit durability due to cell deformation. Optimizing core density and managing delamination issues also remain areas of concern.

In order to achieve effective strength-to-weight ratio, high-strength fibers such as aramid, carbon, and UHMWPE are often used in fiber-reinforced composites to provide multi-hit resistance and efficient energy dissipation [7]. Also, natural fibers such as flax and jute and synthetic fibers combine to create hybrid composites, which offer economical and environmentally beneficial solutions. The natural fibers help absorb energy, while the synthetic fibers increase load resistance [8]. Further improvements in ballistic performance can be achieved by (a) selection of tuned resin, (b) tailoring the fiber orientation and (c) tuning the fiber-matrix bonding through surface treatment. Despite its effectiveness, polymer matrix fiber composites' structural integrity is compromised by matrix cracking, delamination, and fiber pullout. Fiber composites and sandwich constructions frequently have significant manufacturing costs, and fibers like aramid are moisture-sensitive and degrade in harsh environments [20-21].

To improve structural integrity, functionally graded materials (FGMs) enable smooth property transitions between layers, allowing for direction-specific mechanical responses under high-velocity impacts. By maximizing material qualities where they are most required, this gradient design allows the structure to endure a variety of impact circumstances [9]. When combined, these techniques produce lightweight, multipurpose, ballistic-resistant structures that are appropriate for high-impact uses.

Inclusion composites, using fillers like sand, ceramics, or metallic particles within a polymer matrix, are promising for ballistic protection [10]. These inclusions boost hardness and energy dissipation, enhancing resistance to high-velocity impacts. Sand-filled composites, in particular, increase stiffness and hardness at a low cost, though optimizing particle type, size, and distribution remains one of the daunting tasks. The densification of sand within the matrix limits voids, aiding impact resistance. Key factors like sand preparation, segregation, and curing processes significantly influence these composites' effectiveness [11,12].

In this research work, the influence of sand inclusions in two types of polymer matrix compositions forming a composite has been studied. Refinement of the composite fabrication process has been carried out to obtain the necessary mechanical properties with variations in sand inclusion size and weight fraction. To extract these properties, tests such as Uniaxial Tensile, I-Zod Impact, and Shore-D hardness were conducted, with samples prepared by optimizing the fabrication process in each condition. Properties thus extracted are incorporated in decision parameters for an effective graded stacking sequence for ballistic threats. The stacking sequence is arranged to achieve variable mechanical properties in a single structure by leveraging the tailor-ability of PMSCs through sand inclusions and polymer matrix. The layered structure provides an initial, densely graded sand region for impact absorption and abrasive tip deterioration, mitigating localized stresses through increased contact area. A less dense backing region provides additional residual energy absorption and stress transfer, while the final, minimally dense region acts as a cushion by absorbing bullet momentum. As it is difficult to conduct experimental studies for all the particle sizes and volume fractions, a simulation assailed studies have been carried out. DIGIMAT has been used to evaluate the effective properties of Representative Volume Elements (RVE) by varying inclusion size and volume fraction. The thus formed composites exhibit superior interfacial bonding between the matrix and the fillers, preventing delamination under high-velocity impacts. This structural integrity ensures that the layers remain intact, even under extreme stress, thereby enhancing the material's overall ballistic resistance and energy absorption capabilities.

## 2. Experimental
### 2.1 Materials

Epoxy has been chosen as the matrix for its non-toxicity, ease of handling, and room-temperature curing capability [14]. The reactive epoxide rings form a strong network, offering excellent mechanical strength and rigidity. The primary binding agent for initial sand inclusions was ER099 Epoxy Resin with EH150 hardener, valued for its strength, chemical resistance, low shrinkage, non-crystallizing properties, availability, and cost-effectiveness. This combination tolerates temperatures up to 60°C with a long pot life. In the second experiment, LY556 Bisphenol-A-grade epoxy with HY951 hardener was used for its mechanical stability, thermal resistance up to 80°C, and adjustable reactivity [13]. Construction sand, primarily silica, has been selected for its irregular morphology, enhancing adhesion with minimal voids for better and homogenous structural integrity [14]. Its chemical inertness ensures compatibility with the epoxy matrix, and its abrasive nature improves tapped density, aiding in energy dissipation during projectile impact and reducing fragmentation [14,15].

## 2.2 Sand Processing

Two distinct sand processing procedures were developed for composite samples with ER099 and LY556 epoxy resins. For ER099-based samples, sand was air-dried at 30°C for 30-35 hours, effectively removing moisture and organic impurities. It was then sieved using an electric sieve shaker for 15 minutes to obtain specific particle sizes of 0.15 mm, 0.3 mm, 0.425 mm, and 0.6 mm, enhancing packing density and mechanical strength. Finer particles filled voids, while larger particles improved flowability. The fractions were stored in airtight containers to maintain quality **Figure 2 and Figure 3**. The improved procedure for LY556 samples involved additional steps to enhance sand purity and performance. Sun-drying on a polyethylene layer reduced moisture, while remaining organic impurities (e.g., leaves, roots) were manually removed. An isopropyl alcohol wash further purified the sand, followed by a high-temperature treatment at 140°C in a blast furnace for six hours to ensure thorough moisture removal. Milling with the PULVERISETTE 7 achieved uniform particle sizes under 0.1 µm, enhancing packing density and interlocking. Sieving of sand was performed similar to the previous procedure. **Figure 1** and **Figure 2**.

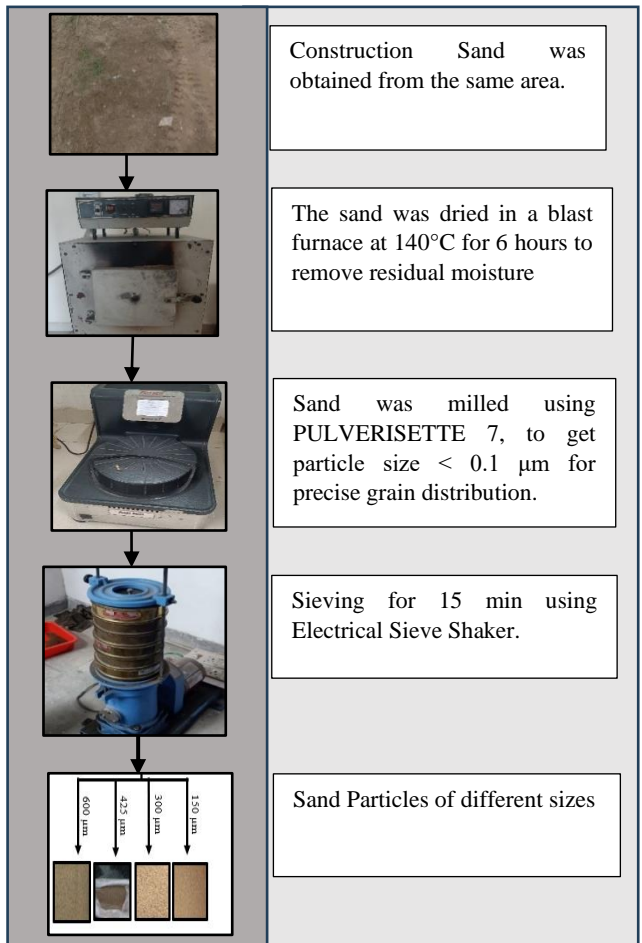

**Figure 1:** Sand Pre-Processing

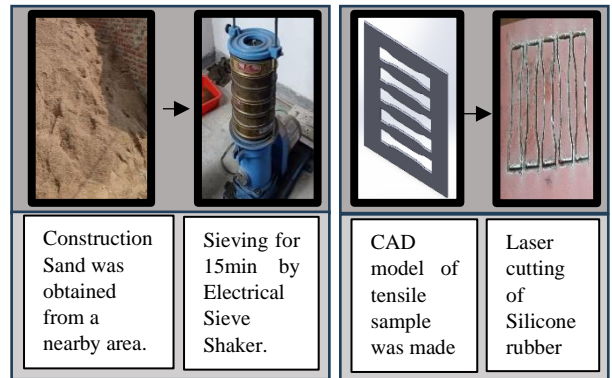

**Figure 2:** Sand Preprocessing and Mold Preparation

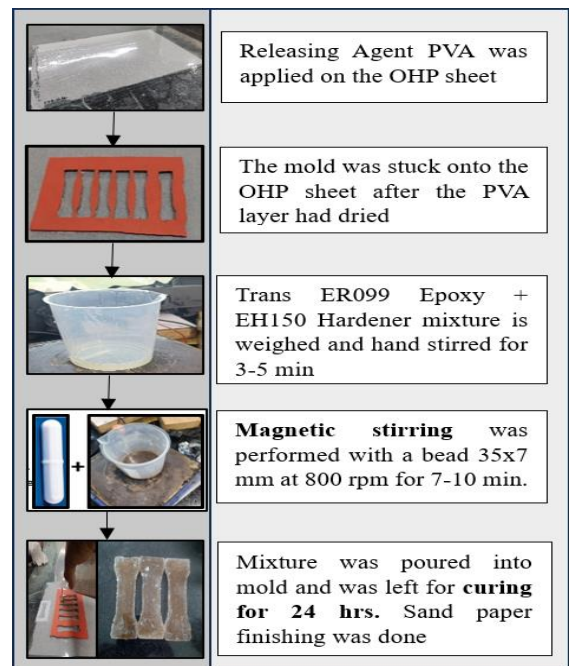

**Figure 3:** Testing Specimen Preparation Process

## 2.3 Mold Preparation

Two mold preparation methods were applied for composite testing: laser cutting and additive manufacturing. For tensile molds, silicone rubber was selected due to its flexibility, durability, and resistance to epoxy resins, avoiding chemical reactions. A 5 mm thick silicone rubber sheet was cut using the BSM Laser Cutting and Engraving Machine, known for its high-speed DSP control and precision, producing reliable tensile molds **Figure 4**.

The improved procedure for LY556 samples involved molds which were created with a 3D printer from Shenzhen Creality 3D Technology Co., Ltd. **Figure 5**. For thermoplastic polyurethane (TPU) molds, a 0.2 mm nozzle, bed temperature of 230°C, nozzle temperature of 70°C, and speed of 65 mm/s were set. TPU's elasticity minimizes shrinkage and simplifies demoulding, ideal for tensile, Izod impact, and Shore-D hardness tests. The molds are reusable, enhancing cost efficiency. For ballistic impact samples, PLA with 90% infill density was used, along with a 0.4 mm nozzle, 200°C nozzle temperature, 60°C bed temperature, and 50 mm/s speed for improved rigidity and thermal stability. PLA molds include clamping features, ensuring secure and accurate testing. These optimized 3D printing parameters and materials improve mold durability, efficiency, and composite testing precision.

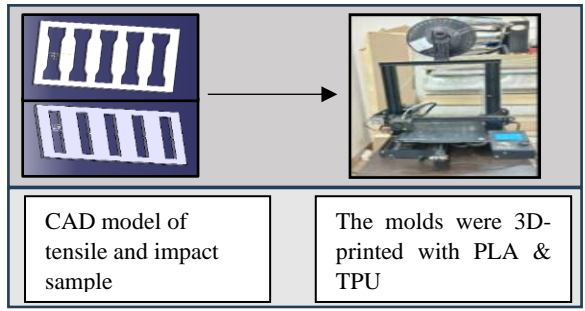

| CAD model of tensile and impact sample | The molds were 3D-printed with PLA & TPU |

**Figure 4:** Mold Preparation

## 2.4 Sample Preparation

Two specimen preparation methods were implemented using ER099 and LY556 epoxies to create durable composite samples with sand filler. For the ER099-based samples, Epoxy trans ER099 spl (Bisphenol-A grade) was weighed using an iScale machine with precise minimum (20/40/100 g) and maximum (10/20/30 kg) weight capacities. Hardener trans EH150 was added at 10% of the epoxy's weight, and sand was weighed to match the target weight fraction in the mixture. For example, in a 100 g epoxy-hardener blend with 20% sand content, 20 g of sand was added. This epoxy-hardener mix was hand-stirred for 3-5 minutes, followed by mixing on a Springer Magnetic Stirrer with Hot Plate at 70% speed for 7-10 minutes to achieve thorough sand dispersion. Molds were lined with an OHP sheet coated in Polyvinyl Alcohol as a releasing agent, ensuring easy removal after curing. The sand-epoxy mixture was poured into the molds and left to air-dry for at least 24 hours. Post-curing, the samples were carefully removed and finished with medium and finer grit sandpaper for a smooth, even surface **Figure 5**. For the LY556-based samples, Epoxy LY556 was combined with Hardener HY951 (Triethylenetetramine) at a 10% hardener ratio. The epoxy and hardener mixture were weighed and hand-stirred for 2-3 minutes and then transferred to a REMI RQ-121 overhead stirrer, running at 800 rpm for 5-6 minutes. Weighted sand was gradually added, extending the mixing time to 20-22 minutes for consistent sand distribution. The molds were prepared similarly, lined with OHP sheets as per section 3.1.3, and the sand-epoxy mix was poured in, air-drying for 24 hours. Post curing of the samples was performed in a GAYATRI SCIENTIFIC WORKS hot-air oven at 100°C for 12 hours to enhance durability and adhesion. After curing, the samples were finished with 120-grit sandpaper and polished on a BAINPOL METCO polishing machine at 120 rpm with 220-grit sandpaper for a refined surface. Each sample was secured on a flat base plate with double-sided tape to ensure stability during finishing. This procedure was replicated across different sand compositions to maintain consistency in testing results.

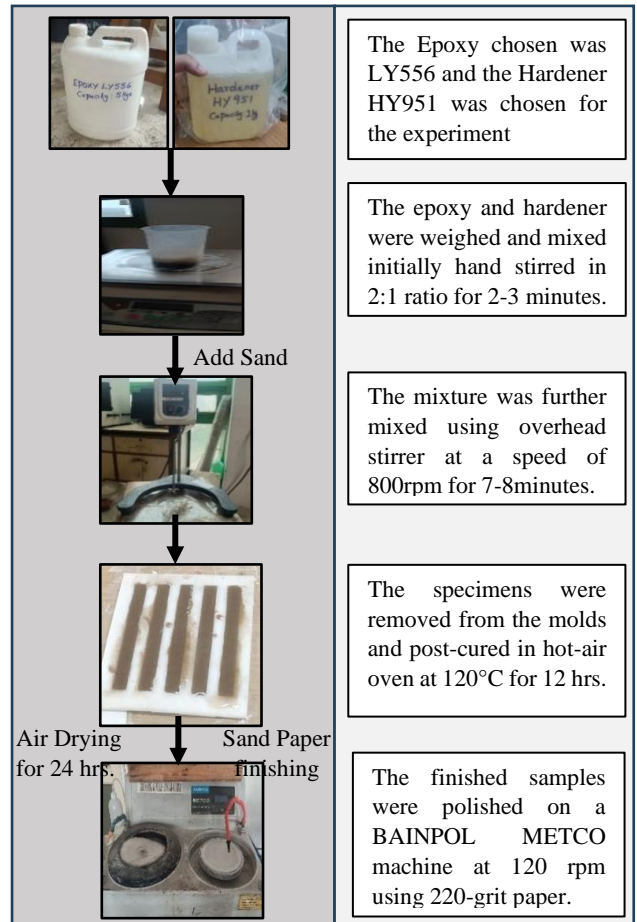

**Figure 5:** Specimen Preparation Procedure

## 2.5 Drawbacks

During sample preparation, issues arose with sand particle settling [15] and weak bonding. High viscosity leads to reduced voids and improved bonding. Therefore, the low viscosity of Trans ER-099 epoxy might have been the issue. Additional challenges included uneven silicone mold finishing and bubbles from magnetic stirring. The lack of post-curing weakened the samples' mechanical strength. Prior studies [14] suggest that post-curing at optimal temperatures can enhance bonding, toughness, and strength, highlighting a key area for improvement

## 2.6 Tensile Test

Tensile Strength is a characteristic that defines the ability of the material to resist breaking under tension. This property is influenced by the molecular structure of the epoxy resin and the nature of the reinforcing materials, such as sand inclusions. In epoxy-based composites, tensile strength directly correlates with the crosslink density of the cured resin, the quality of the bonding between the epoxy matrix and the reinforcement [14]. The tensile tests were performed on both neat epoxy samples with different sizes and weight fractions of sand inclusions in accordance with the ASTM Standard D638. The testing was performed using Zwick/Roell Z010. A load cell of 10kN was used and the testing was performed corresponding to the speed 5 mm min$^{-1}$. The specimens measured 19mm in width, 50mm in gage length and the distance between the grips was 115mm.

## 2.7 Impact Test

Impact Strength is a characteristic that defines the ability of the material to resist sudden forces or shock by determining the energy absorbed during the fracture. Izod Impact test was performed on both neat epoxy samples with different sizes and weight fractions of sand inclusions in accordance with the ASTM Standard D256-23. The specimens measured 63.5 mm in length, 12.7 mm in width, and 5 mm in thickness. A V-notch of angle 67.5 degree. was given at the center of the specimen. Each test specimen was placed atop a vertical cantilever beam and fractured by a single pendulum striking the center of the sample.

## 2.8 Shore-D Hardness Test

Hardness is the measure of the material's resistance to plastic deformation by indentation. The hardness of both neat epoxy samples with different sizes and weight fractions of sand inclusions was measured using Shore D durometer in accordance with ASTM standard D2240-15. The specimens are of dimensions 35mm in length ,25 in width and 6mm in thickness.

# 3. Effective Properties Evaluation

In order to analyze the effect of inclusion size and volume fraction on the macro-scale effective properties were extracted from the RVE which was done using commercial FEA tool DIGIMAT FEA. Spherical inclusions with a random but uniform distribution over the matrix were considered. The variation has been carried out with respect to inclusion size and volume fraction. Displacement boundary condition has been imposed as negligible variance in effective properties has been imposed by other boundary conditions. Effective Properties are extracted by Volume Averaging Method where the stresses and strains in the respective directions as per the boundary conditions imposed are averaged out over the volume which ultimately gave effective properties in the particular direction and plane. Some typical configurations and distributions are as shown in **Figure 6.** the isotropic properties assigned to the constituents i.e. epoxy and sand were as in [13,16].

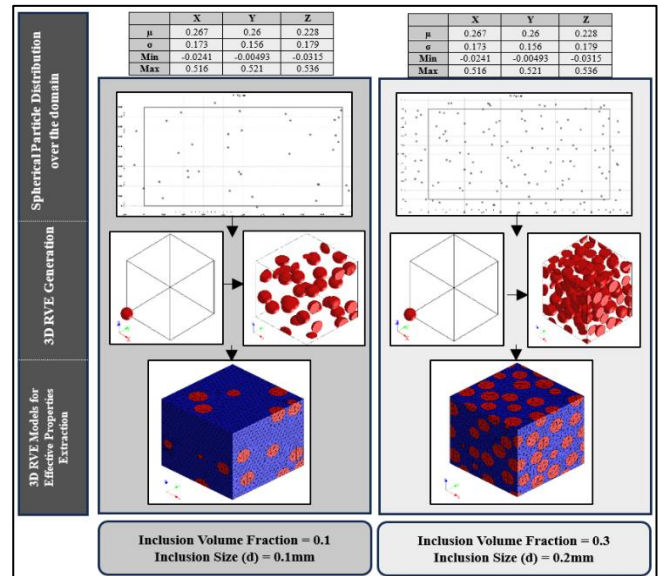

**Figure 6**: Modelling of RVE for Varying inclusion size

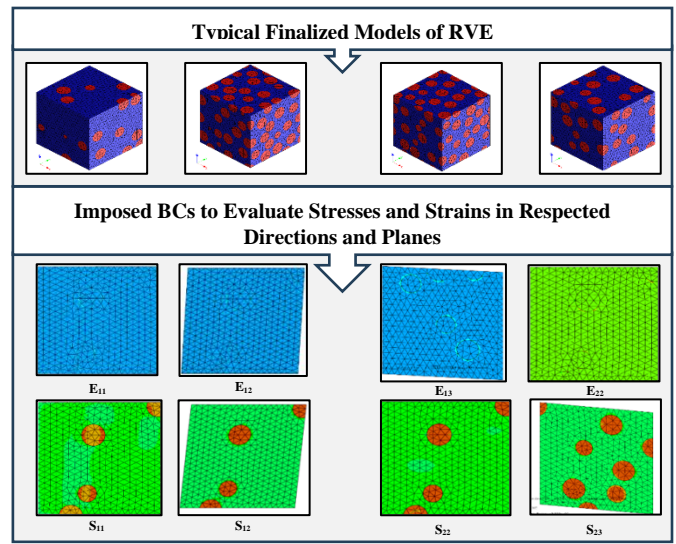

**Figure 7:** BCs applied to extract effective properties by varying inclusion size and volume fraction

Effective properties were extracted by imposing appropriate boundary conditions on varied RVEs where the homogenous isotropic properties for spherical sand inclusions are considered as in [16]. A schematic of the different BCs applied on the RVE in order to extract effective properties in respective directions and planes is shown in **Figure 7.** The simulation study reveals that through this approach, a comprehensive understanding of how these parameters influence critical material properties was developed.

# 4. Results and Discussion

## 4.1 Microstructural Analysis

Optical images generated using an Olympus base microscope were classified based on sand inclusion size and volume fraction. The refined manufacturing process enabled a more uniform sand distribution up to a certain limit, with fewer agglomerations in samples with smaller grain sizes and lower volume fractions, as observed in **Figure 8 a)**. Although larger particle sizes did not significantly increase agglomerations, higher volume fractions led to unevenly dense regions due to particle clustering, as shown in **Figure 8 d)**. Voids were generally observed across samples as shown in **Figure 8 c)**, though they were minimal and likely resulted from minor inconsistencies in filling or dispersion during manufacturing. Under load, cracks formed in the less dense, neat epoxy areas, which were structurally weaker than the sand-reinforced regions as shown in **Figure 8 e)** and **Figure 8 f)**.

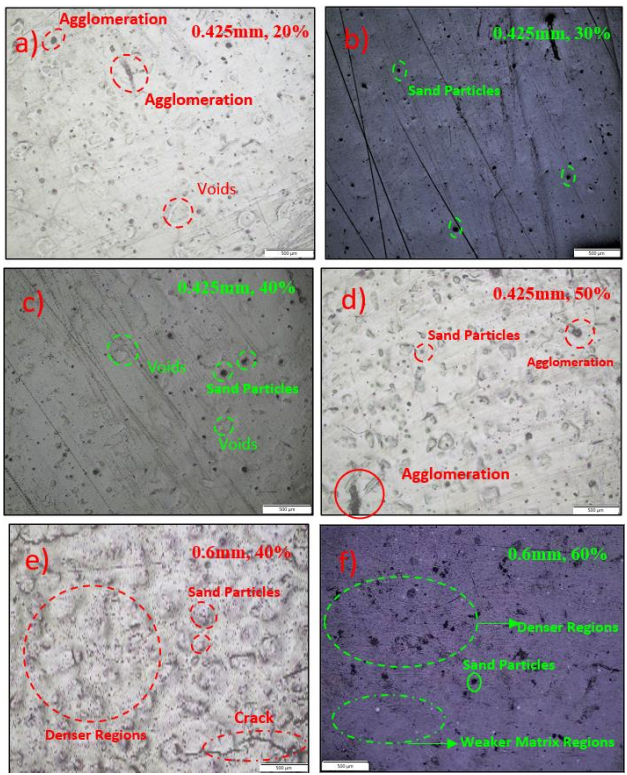

**Figure 8:** Images Captured through Optical Microscope of PMSCs

## 4.2 Effective Properties Outputs

The Effective properties for Varied RVEs were extracted where variation in the Inclusion size from 0.1 to 1 mm in diameter and the volume fraction from 0.2 to 0.8 in most of the simulations were carried out.

The simulation results for RVE clearly demonstrate a significant increase in the elastic moduli in the three principal directions of the PMSCs $E_1$, $E_2$, and $E_3$, as both the inclusion diameter and volume fraction were increased as depicted in **Figure 9 a), b), c).** The Properties showed minimal variation in the principal directions demonstrating a quasi-isotropic Nature of PMSCs**.** An increase in Moduli results from larger inclusions contributing greater stiffness, effectively restricting matrix deformation. With increased size or volume, inclusions act as rigid supports, enhancing the composite's overall stiffness. On the contrary, Poisson's ratios ($v_{21}$, $v_{13}$, etc.) decrease with larger inclusion size and volume fraction, as stiffer inclusions limit lateral deformation, aligning with increased PMSCs rigidity. Additionally, shear moduli $G_{12}$, $G_{23}$, $G_{13}$ increase due to inclusions effectively resisting shear by evenly transferring loads **Figure 9 d), e), f)**.

In conclusion, the study effectively captures the influence of inclusion size and volume fraction on the mechanical properties of the RVE. The results underline the critical role of these parameters in enhancing the stiffness and shear moduli while reducing lateral deformation, contributing to the material's overall robustness. This approach of varying inclusion size and volume fraction can serve as a powerful design tool in optimizing composite materials for high-performance applications.

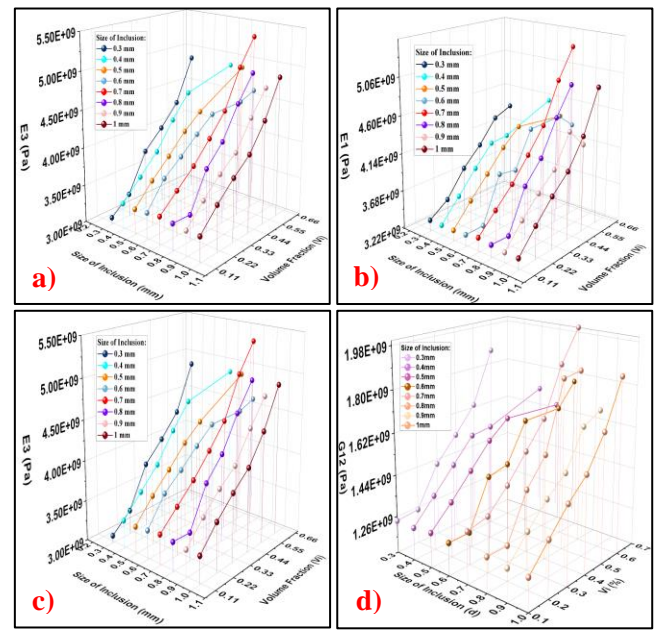

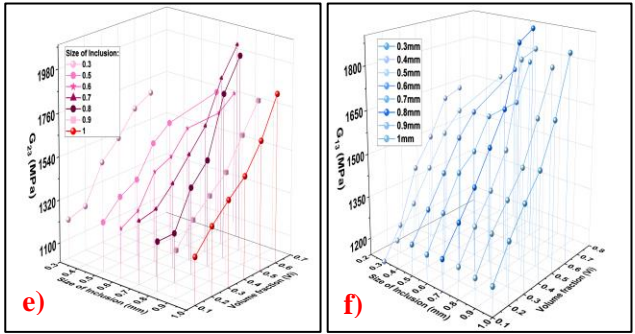

**Figure 9:** Effective properties in respective directions and planes

### 4.3 Experimental Material Properties

Experimental testing including uni-axial tensile test, Izod impact test, and Shore-D hardness test was performed adopting a refined manufacturing process as discussed in **Section 2** with appropriate standard specimen sizes and dimensions specified in **Sections 2.6**, **2.7**, **2.8.** The major comparison was made between two grades of polymer matrixes and refinement has been carried out in each iteration in order to obtain the consistency of obtained results and thereby select the best suitable grade with the optimized manufacturing process.

Tensile testing of ER 099 (Epoxy) with EH 150 (Hardener) and 0.3 mm sand particles, varying sand volume fractions from 0.30 to 0.50, revealed that a 30% sand volume increased the modulus by 50.20% to 366 MPa compared to neat epoxy. At 40% and 50% volumes, modulus reached 534.67 MPa and 539 MPa, marking 119% and 121.20% increases, respectively as shown in **Figure 10 a)**. Impact strength for the same configuration rose to 110 J/m and 135 J/m at 40% and 50% sand volume, showing 7% and 32% improvements over the neat mixture represented in **Figure 10 b)**. While higher sand volume fractions presented challenges due to manufacturing irregularities discussed in **Section 2.5**, these insights highlighted areas for improvement, particularly in mitigating agglomeration to reduce localized stress in weaker matrix regions.

A more viscous, harder epoxy grade **Section 2.1** was used, offering advantages over the previous grade. Tensile tests with 0.3 mm and 0.425 mm sand particles at increasing volume fractions showed property enhancement. For 0.3 mm sand, Young's Modulus increased by 51% (1810 MPa) at 70% volume, while for 0.425 mm sand, it rose to 1445 MPa i.e. 25% increase at 60% volume, as shown in **Figure 10 e)**. Higher sand volumes up to 70% for 0.425mm sand size degraded modulus, suggesting manufacturing improvements. Impact strength reached 210 J/m for 0.3 mm sand at 70% volume and showed gradual increments for 0.45 mm and 0.6 mm sand 100 J/m and 150 J/m at 60% and 70% volumes

**Figure 10 c)**. Shore-D hardness tests indicated values of 83.5, 89, and 92.5 for 0.3 mm, 0.425 mm, and 0.6 mm sand at 70% volume Figure, with a 13.7% increase over the pure matrix **Figure 10 d)**. The Comparison between the simulation and Experimental Properties extracted are as in **Figure 10 f)**.

The use of the Second Grade of Epoxy Hardener Mixture significantly improved the performance of the PMSCs, with results aligning closely with findings in relevant literature [11,12] and the homogenized properties extracted in **Section 3.2**. This improvement in the manufacturing procedure demonstrates a viable solution for ballistic applications, offering lightweight, cost-effective, and readily available materials for ballistic structures.

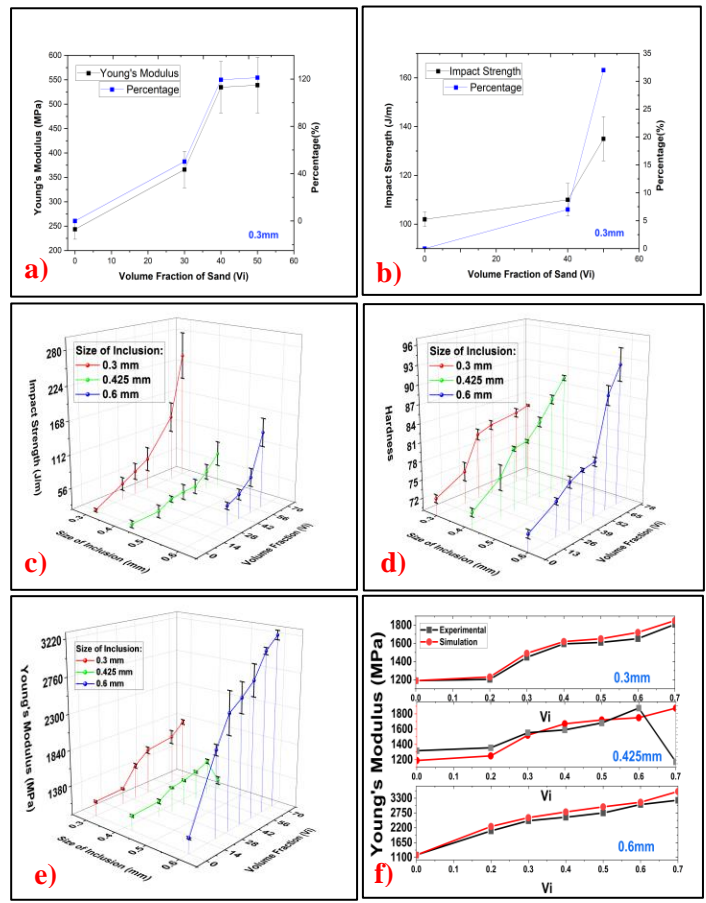

**Figure 10:** Experimental Properties

### 4.4 Ballistic Samples Assertions

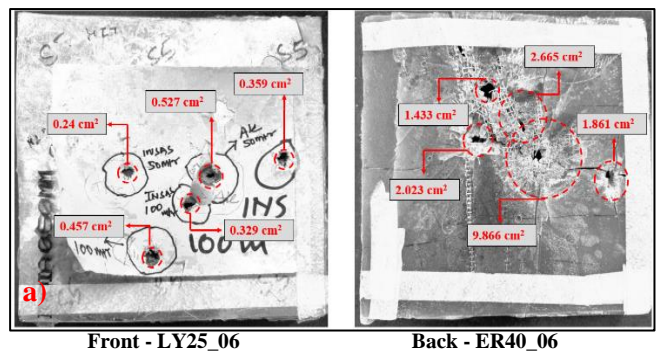

**Front - LY25_06**  **Back - ER40_06**

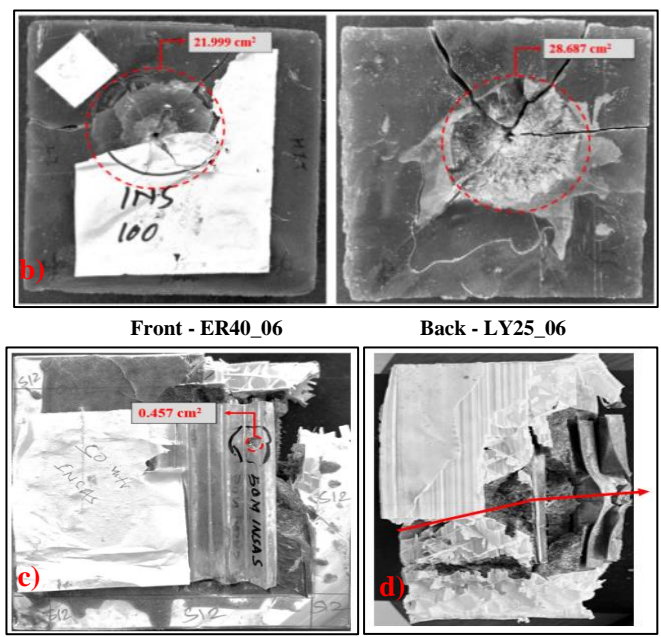

**Figure 11:** Ballistic Testing on Configured Samples

The testing setup apparatus was with different firing distances along with the guns used to perform testing, with sample configurations detailed in [17]. Corrugated sheets and Mesh were used to act as a stiff structure along with a potent of mesh to further distribute stresses over larger area. Both samples i.e. **Figures 11 a)**, **b)** underwent ballistic testing with an INSAS rifle from 100 meters. Comparative analysis shows a larger bullet impact area in sample LY25_06 than in ER40_06, indicating greater bullet deformation upon impact with LY25_06, likely due to the higher stiffness of the LY556 epoxy. The stronger polymer-inclusion bonding in ER40_06 contributed to enhanced material integrity, as evidenced by more visible cracks in LY25_06, suggesting it absorbed a larger share of impact energy. **Figure 11 c)** configures as in [17], tested ballistically with an INSAS rifle at 50 meters. increased layer counts introduced complexity, leading to delamination due to stress concentration at interfaces. **Figure 11 d)** shows sample LY80_04_1 [17], tested with a SIG rifle at 50 meters. Bullet trajectory deviation was observed, likely due to varying properties across layers, absorbing and redirecting impact energy. However, delamination occurred in both samples due to interface stress concentration, weakening bonds despite overall material performance [18].

## 5. Conclusion

This work presents a detailed study on developing a novel sand-based functionally graded composite optimized for ballistic resistance. The fabrication process and experimentally verified homogenized material properties are discussed, followed by a comprehensive ballistic analysis. Key observations include:

1) Mechanical properties improved significantly, with modulus increasing up to 125% and impact strength by 32% with increase in volume fraction of sand as observed in **Section 4.3**

2) With an increase in inclusion size it has been observed that the stiffness had increased, restricting matrix deformation and decreasing Poisson's ratio, contributing mainly to lateral deformation. Both elastic and shear moduli also increased as reported in **Section 4.2**.

3) LY-556 ballistic samples showed greater impact areas due to higher stiffness, improving impact absorption compared to ER-099. However, increased layering led to delamination from stress concentration at interfaces as observed in **Section 4.4**.

4) One of the critical observations is that Gradation Proved Superior to uniform distribution, reducing bullet's kinetic energy over shorter intervals and decreasing penetration depth as observed in **Section 4.4**.

Further analyses are being carried out to improve and optimize the mechanical properties which will be communicated in the future articles.